\documentclass[%
 preprint,prl,
 amsmath,amssymb,
 aps,superscriptaddress
]{revtex4-1}

\usepackage{graphicx}% Include figure files
\usepackage{bm}% bold math
\usepackage{subfig}
\usepackage{float}

\usepackage{color}
\usepackage[normalem]{ulem}

\begin{document}

\preprint{APS/123-QED}

\title{Post-$GW$ theory and its application to pseudogap in strongly correlated  system}

\author{Hui Li}
 \email{physicslihui@zju.edu.cn}
\thanks{These authors contributed equally to this work.} 
\affiliation{%
Institute for Advanced Study in Physics and School of Physics, Zhejiang University, Hangzhou 310027, China
} 

\author{Yingze Su}%
\thanks{These authors contributed equally to this work.} 
\affiliation{%
School of Physics, Peking University, Beijing 100871, China
}

\author{Junnian Xiong}
\affiliation{%
School of Physics, Peking University, Beijing 100871, China
}

\author{Haiqing Lin}%
 \email{hqlin@zju.edu.cn}
\affiliation{%
Institute for Advanced Study in Physics and School of Physics, Zhejiang University, Hangzhou 310027, China
} 

\author{Huaqing Huang}%
 \email{huaqing.huang@pku.edu.cn}
\affiliation{%
School of Physics, Peking University, Beijing 100871, China
}

\author{Dingping Li}%
 \email{lidp@pku.edu.cn}
\affiliation{%
School of Physics, Peking University, Beijing 100871, China
}

\date{\today}

\begin{abstract}
The $GW$ approximation is a widely used framework for studying correlated materials, but it struggles with certain limitations, such as its inability to explain pseudogap phenomena. To overcome these problems, we propose a systematic theoretical framework for Green's function corrections and apply it specifically to the $GW$ approximation. In this new theory, the screened potential is reconnected to the physical response function, i.e. the covariant response function proposed in \cite{cGW_2023}, rather than using the RPA formula. We apply our scheme to calculate Green's function, the spectral function, and the charge compressibility in the two-dimensional Hubbard model. Our scheme yields significant qualitative and quantitative improvements over the standard $GW$ method and successfully captures the pseudogap behavior.
\end{abstract}

\maketitle

%\tableofcontents

\textit{Introduction.}---
 Green's function determines the single-particle spectrum, band structure, energy, and other important properties. However, accurately calculating Green's function in strongly correlated systems is one of the greatest challenges and crucial problems in condensed matter physics. One commonly used approach to evaluate Green’s function is the approximation by truncation of Dyson's equation\cite{altland_simons_2010,coleman_2015}. In particular, the $GW$ approximation is one of the most widely used approaches, which contains the screened effect in its self-energy, and it has been successfully applied in metals, semimetals, nanomaterials and so on\cite{Hedin, Yuchen2016, Aryasetiawan_1998, GW_review_2019, Patrick2019}. However, it also suffers from some serious shortcomings, including the absence of satellite peaks in the spectrum, and the inability to describe the pseudogap in strongly correlated systems\cite{Lucia2018,ZhipengSun2021}. 

The pseudogap, characterized by the depletion of electronic states in the normal state near the Fermi surface, is a phenomenon observed in many strongly correlated systems, including copper-oxide superconductors and unitary Fermi gases.\cite{pseudogap1999,pseudogap2001,JinD2008pseudogap,haussmann2009spectral,pseudogap2009,JinD2010pseudogap,pseudogap2024}. Understanding the pseudogap is widely considered crucial for unraveling the microscopic mechanisms underlying high-temperature superconductivity.
To address the limitations %issues 
of the $GW$ approximation, various extensions have been proposed, such as combing it with the dynamical mean-field theory (DMFT) \cite{choi2016} or extending to higher-order vertex approximations in Hedin's equation \cite{takada2001inclusion,vertex_2007,takada2011,vertex_2012,vertex_2014,vertex_2015,kutepov2016,ren2015,kutepov2017one,vertex_2020,kutepov2022,vertex_2023}. Nevertheless, these attempts have yet to fully resolve the issues related to the pseudogap and other strongly correlated phenomena within the $GW$ framework. 

In this Letter, we propose a general framework to improve Green's function of the existing many-body approximation theories, which we term the post theory. We apply this scheme to the $GW$ approximation to obtain the post-$GW$ theory, a significant advancement over the conventional $GW$ theory. Our motivation arises from the violation of the intrinsic relationships for the high-order correlation functions in the original theories due to truncation, for example, the relation between the screened potential and the charge/spin correlation. In the $GW$ theory, such a relation is violated due to the vertex approximation. In Ref. \cite{cGW_2023}, we introduce the covariant theory to obtain physical charge/spin correlation functions, which preserve the fluctuation-dissipation theorem (FDT) \cite{Kubo_1966} and the Ward-Takahashi Identity (WTI) \cite{book-peskin}. In our post-$GW$ theory, we replace the screened potential in the $GW$ Green's function with the physical one, which is determined by the covariant response function. 

To validate the post framework, we apply the post-$GW$ approach to 2D Hubbard models to calculate Green's functions and compare our results with those obtained from the Determinantal Quantum Monte Carlo (DQMC) method. Moreover, we use the Nevanlinna analytical continuation to calculate the spectral function, demonstrating that
the post-$GW$ approach can indeed describe the pseudogap. Finally, we calculate the compressibility $\partial n/\partial\mu$ and compare it with DQMC results from Ref.~\cite{Huang_2019}.
%more sentence

\textit{General formalism.}
---For an arbitrary exact many-body theory, it can always be expressed as:
\begin{align}
G&=F_0(G,L_1,\cdots,L_n,\cdots),\label{eq:postcov-exactG}\\
L_n&=F_n(G,L_1,\cdots,L_n,\cdots)\nonumber.
\end{align}
where $G$ is the Green's function, $L_n\equiv\frac{\delta^n G}{\delta \phi^n}$ is the n-th functional derivative to an external source $\phi$.
Functional $F_0$ presents as Dyson's equation $G^{-1}=G^{-1}_0-\Sigma[G]$ \cite{cubic2018,Fan2020,Sun2021}.
 However, such equations cannot be exactly solved for most correlated systems because they are not closed (infinite hierarchy). Hence, we need approximations to make these equations solvable (for example, $L_{k}=0$ for all $k>n$). The approximate equations take the form after taking $\phi\rightarrow0$ (named on-shell equations):
\begin{equation}
    \left\{\begin{matrix}
G^{\mathrm{tr}}=\tilde{F}_0(G^{\mathrm{tr}},L_1^{\mathrm{tr}},\cdots,L^{\mathrm{tr}}_n),\\
 L_1^{\mathrm{tr}}=\tilde{F}_1(G^{\mathrm{tr}},L_1^{\mathrm{tr}},\cdots,L^{\mathrm{tr}}_n),\\
 \cdots\\
L_n^{\mathrm{tr}}=\tilde{F}_n(G^{\mathrm{tr}},L_1^{\mathrm{tr}},\cdots,L^{\mathrm{tr}}_n).

\end{matrix}\right.\label{eq:postcov-app}
\end{equation}
Eqs.~(\ref{eq:postcov-app})  are closed and can be solved.  The formations for $\tilde F$ may differ from $F$ in the exact theory. It should be pointed out that, the solutions of Eqs.~(\ref{eq:postcov-app}) are ``truncated" correlations, i.e. they are not ``physical" because these correlations violate the definition $L_k=\frac{\delta^k G}{\delta \phi^k}$. This feature leads to the ``truncated" Green's function being unable to obey some crucial relations for cumulants $L_k$, for example, WTI and FDT \cite{cGW_2023}. Now we propose the following formalism to calculate the physical correlations.

Next, we consider the Eqs.~(\ref{eq:postcov-app}) with the source term $\phi\neq0$, referred to as the off-shell equations. To obtain the physical correlation,  we calculate the functional derivative with respect to $\phi$ in off-shell Eqs.~(\ref{eq:postcov-app}) :
\begin{equation}
    \left\{\begin{matrix}
\dot G^{\mathrm{tr}}(\phi)=\frac{\delta\tilde F_0}{\delta G^{\mathrm{tr}}}\dot G^{\mathrm{tr}}(\phi)+\sum_{k=1}^{n}\frac{\delta\tilde F_0}{\delta L_k^{\mathrm{tr}}} \dot L_k^{\mathrm{tr}}(\phi)
,\\
\dot L_1^{\mathrm{tr}}(\phi)=\frac{\delta\tilde F_1}{\delta G^{\mathrm{tr}}}\dot G^{\mathrm{tr}}(\phi)+\sum_{k=1}^{n}\frac{\delta\tilde F_1}{\delta L_k^{\mathrm{tr}}}\dot L_k^{\mathrm{tr}}(\phi),\\
 \cdots\\
\dot L_n^{\mathrm{tr}}(\phi)=\frac{\delta\tilde F_n}{\delta G^{\mathrm{tr}}}\dot G^{\mathrm{tr}}(\phi)+\sum_{k=1}^{n}\frac{\delta\tilde F_n}{\delta L_k^{\mathrm{tr}}}\dot L_k^{\mathrm{tr}}(\phi).\\
\end{matrix}\right.\label{eq:postcov-covarinat}
\end{equation}
Here we denote $\dot G^{\mathrm{tr}}=\frac{\delta G^{\mathrm{tr}}}{\delta\phi}\equiv L_1^{cov}$,  $\dot L_k^{\mathrm{tr}}=\frac{\delta L^{\mathrm{tr}}_k}{\delta\phi}$ and call Eqs.~(\ref{eq:postcov-covarinat}) covariant equations, which are linear for $L_1^{cov}$ and $\dot L_k^{\mathrm{tr}}$.  The Eqs.~(\ref{eq:postcov-covarinat}) can be solved by taking the shell limit $\phi\rightarrow 0$.

Similarly, the second-order functional derivative provides a set of linear equations about $\ddot G^{\mathrm{tr}}$, $\ddot L_k^{\mathrm{tr}}$, and one can evaluate $\ddot G^{\mathrm{tr}}=\frac{\delta^2 G^{\mathrm{tr}}}{\delta\phi^2}\equiv L_2^{cov}$. We can repeat this procedure to obtain all physical correlations $\frac{\delta^n G^{\mathrm{tr}}}{\delta\phi^n}\equiv L_k^{cov}$. We should point out that the solutions of covariant equations are physical because they are defined through the functional derivative, satisfying Kubo's formula automatically \cite{Kubo_1966} (relation between the response function and the correlation function). 

The new Green's function is proposed by replacing the $L_k^{\mathrm{tr}}$ in Eqs.~(\ref{eq:postcov-app}) by the physical correlation $L_k^{cov}$:
\begin{equation}
    G_{\mathrm{post}}=\tilde{F}_0(G^{\mathrm{tr}},L_1^{cov},\cdots,L_n^{cov}).\label{eq:postcov-postG}
\end{equation}
The method to obtain $G_{\mathrm{post}}$ presented above is called post theory for the approximation of Eqs.~(\ref{eq:postcov-app}) in this article. The validity of the post framework can be verified by applying it to exactly solvable toy models (see
Supplemental Material (SM) for detailed calculations). Subsequently, we will consider its application to the $GW$ approximation.

\textit{Hedin's equations and $GW$ equations for general action.}---We start with a general Matsubara action form at finite temperature \cite{Aryasetiawan2008}:
\begin{equation}
\begin{aligned}
    S[\psi^*,\psi]
    =&-\sum_{\alpha_1\alpha_2}\int\mathrm d(12)\ 
    \psi_{\alpha_1}^*(1)T_{\alpha_1\alpha_2}(1,2)\psi_{\alpha_2}(2)\nonumber\\
    &+\frac{1}{2}\int\mathrm d(12)\ 
    \sum_{ab}\sigma^a(1)V^{ab}(1,2)\sigma^b(2),
    \end{aligned}
    \label{S}
\end{equation}
where $\psi,\psi^*$ are Grassmannian fields,  $\sigma^a(1)=\sum_{\alpha\alpha'}\psi^*_{\alpha}(1)\tau^a_{\alpha\alpha'}\psi_{\alpha'}(1)$ are charge/spin operators and  $\tau^a(a=0,x,y,z)$ are Pauli matrices. The Greek letters like $\alpha$ indicate the spin projection. The label $(1)=(x_1,\tau_1)$ is a generalized coordinate, containing the space coordinate $x_1$, and the imaginary time coordinate $0\leq\tau_1< \beta$, where $\beta=1/(k_BT)$ and $k_B,T$ are the Boltzmann constant and thermal-dynamic temperature respectively. The symbol $\int\mathrm d(1)$ represents an integral over all space and time coordinates for a continuous system or a summation over all lattice and time coordinates for a discrete system. The two-body interaction is real symmetric, i.e., $V^{ab}(1,2)=V^{ba}(2,1)$.
We introduce an external vector source $\vec{\phi}\left(1\right)$ coupled to the charge/spin operator $\vec{\sigma}\left(1\right)$, and obtain
the perturbed action
\begin{equation}
S\left[\psi^{\ast},\psi;\vec{J}\right]=S\left[\psi^{\ast},\psi\right]
-\sum_{a}\int \mathrm d\left(1\right)\ 
\phi^{a}\left(1\right)\sigma^{a}\left(1\right).
\label{eq:perturbed}
\end{equation}
By virtue of the grand partition function $Z=\int\mathcal{D}\left[\psi^{\ast},\psi\right]\ \text{e}^{-S\left[\psi^{\ast},\psi;\vec{\phi}\right]}$,
one can define the one-body Green's functions as
\begin{equation}
    G_{\alpha_1\alpha_2}(1,2) = \left\langle \psi^*_{\alpha_2}(2)\psi_{\alpha_1}(1) \right\rangle.
\end{equation}
$\left< \cdots \right>$ presents for $\frac{1}{Z}\int \mathcal{D}[\psi^*,\psi]\cdots e^{-S}$, and $\mathcal{D}[\psi^*,\psi]=\mathcal{D}\psi^*\mathcal{D}\psi$ defines the functional integral measure. Due to the spin structure of the interaction, we denote the matrix in the spin space as
\begin{equation}
    \underline{X}=
    \begin{bmatrix}
    X_{\uparrow\uparrow}& X_{\uparrow\downarrow}\\
    X_{\downarrow\uparrow}&  X_{\downarrow\downarrow}
    \end{bmatrix},
\end{equation}
and its trace is denoted by $\mathrm{Tr}[\underline{X}]= X_{\uparrow\uparrow}+X_{\downarrow\downarrow}$.
Then one can derive the generalized Hedin’s equations for the action Eq.(\ref{S}):
\begin{align}
 &\underline{G}^{-1}\left(1,2\right) 
     =\underline{G_0}^{-1}\left(1,2\right)-\underline{\Sigma}\left(1,2\right),   \label{eq:DS-G}\\
    &\underline{\Sigma}\left(1,2\right) 
     =-\sum_{ab}\int\mathrm d(34)\ 
    \underline{\tau}^{a}\underline{G}\left(1,4\right)\underline{\Lambda}^{b}(4,2;3)W^{ba}\left(3,1\right).  \nonumber 
\end{align}
where $G_0$ is the Hartree propagator, which takes the form $\underline{G_0}^{-1}(1,2)=\underline T(1,2)+\delta\left(1,2\right)\sum_{a}\underline \tau^{a}v^{a}\left(1\right)$.
And $ v^{a}\left(1\right)\equiv \phi^{a}\left(1\right)-\sum_{c}\int\mathrm d\left(3\right)\ V^{ac}\left(1,3\right)\left\langle \sigma^c(3) \right\rangle$ is the effective potential. Here the vertex is obtained by $\underline\Lambda^a(1,2;3)=\delta\underline G^{-1}(1,2)/\delta v^a(3)$. The screened potential here is defined by $W^{ab}(1,2)=\sum_c\int\mathrm d(3)\ \frac{\delta v^a(1)}{\delta \phi^c(3)}V^{cb}(3,2)$, and it can be simplified by introducing the polarization function $P^{ab}(1,2)=\int d(34)\mathrm{Tr}[\underline\tau^a\underline G(1,3)\underline\Lambda^b(3,4,2)\underline G(4,1)]$ as:
\begin{equation}
(W^{-1})^{ab}(1,2)=(V^{-1})^{ab}(1,2)-P^{ab}(1,2).\label{eq:DS-P}
\end{equation}
Hedin's equations consist of Eqs.~(\ref{eq:DS-G},\ref{eq:DS-P}). It should be noted that there exists an exact relation between the charge/spin correlation $\chi^{ab}(1,2)=\delta\left\langle \sigma^a(1) \right\rangle/\delta \phi^b(2)=\left\langle \sigma^a(1)\sigma^b(2) \right\rangle-\left\langle \sigma^a(1) \right\rangle\left\langle \sigma^b(2) \right\rangle$ and the screened potential
 \begin{equation}
\begin{aligned}
&W^{ab}(1,2)=V^{ab}(1,2)\\
&-\sum_{cd}\int\mathrm d(34)\ V^{ad}(1,4)\chi^{cd}(4,3)V^{cb}(3,2),
\end{aligned}
\label{eq:W-chi}
\end{equation}
where we use the definitions of $W,v,\chi^{ab}$.

The simplest approximation for the vertex function $\underline\Lambda^a(1,2;3)\approx\underline{\tau}^a\delta(1,2)\delta(1,3)$ results in the $GW$ approximation with the following self-energy and polarization function:
\begin{align}
     \underline{\Sigma}^{\mathrm{tr}}\left(1,2\right)  =&-\sum_{ab} \underline{\tau}^{a}\underline{G}^{\mathrm{tr}}\left(1,2\right)\underline{\tau}^{b}W^{ba}_{\mathrm{tr}}\left(2,1\right),\label{eq:ggw-sigma}\\   P^{ab}_{\mathrm{tr}}\left(1,2\right)=&\text{Tr}\left[\underline{\tau}^{a}\underline{G}^{\mathrm{tr}}\left(1,2\right)\underline\tau^b\underline{G}^{\mathrm{tr}}\left(2,1\right)\right].\label{eq:ggw-p}
\end{align}
One can solve the on-shell $GW$ equations ($\vec\phi=0$) to obtain the truncated Green's function $G^{\mathrm{tr}}$ and screened potential $W_{\mathrm{tr}}$. 
 
It should be noted that according to Eq.~(\ref{eq:W-chi}), the screened potential $W^{ab}$ is induced by the charge/spin correlation $\chi^{ab}$ in the exact theory. However, the relation (\ref{eq:W-chi}) is broken by the vertex approximation in the $GW$ theory. By combining Eq.~(\ref{eq:W-chi}) and Eq.~(\ref{eq:ggw-p}), we can find the screened potential $W_{\mathrm{tr}}$ in the $GW$ equations is induced by the RPA correlation $\chi_{\mathrm{RPA}}=P_{\mathrm{tr}}/(VP_{\mathrm{tr}}-1)$:
 \begin{equation}
\begin{aligned}
W_{\mathrm{RPA}}=V
-V\chi_{\mathrm{RPA}}V.\label{eq:GW-RPAW}
\end{aligned}
\end{equation}
$\chi_{\mathrm{RPA}}$ is unphysical since it violates the basic definition $\chi^{ab}(1,2)=\delta\left\langle \sigma^a(1) \right\rangle/\delta \phi^b(2)$ , therefore violates the WTI and the FDT. So our goal is to reconnect the screened potential to the physical correlation function in the $GW$ framework.

\begin{figure*}
\includegraphics{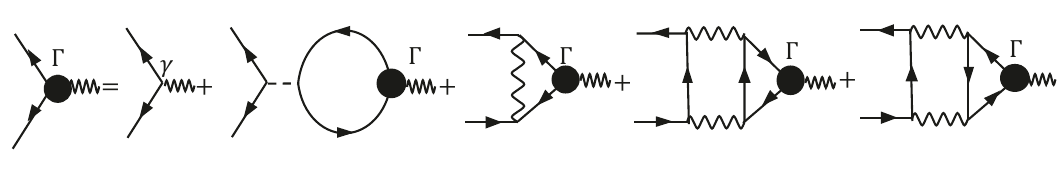}
\captionsetup{justification=raggedright, singlelinecheck=false}
\caption{\label{FDvertex} The Feynman diagram of the full $cGW$ vertex function for translational invariant systems in the momentum space. The dashed line denotes the interaction, the wave line denotes the $W_{\mathrm{tr}}$, and the solid line represents the $G_{\mathrm{tr}}$ in the $GW$ equations.}
\end{figure*}

\textit{Post-$GW$ approach.}---In Ref.\cite{cGW_2023}, we proposed the covariant $GW$ (cGW) approach to calculate charge/spin correlation functions $\chi^{ab}_{cov}=\frac{\delta\left<  \sigma^a\right>}{\delta\phi^b}$ with the Feynman diagrams as shown in Fig.~\ref{FDvertex}, which can both preserve the FDT and the WTI within the $GW$ framework. In the covariant scheme, all correlation functions should be defined as the response of the physical quantity in the presence of an external potential in each many-body theory. It is consistent with the approach for calculating the high-order correlation $L_k^{cov}$ in the post framework. 

Now we calculate the post Green's function by using the covariant charge/spin correlation $\chi^{ab}_{cov}$ and the physical screened potential, which is the application of the post framework above in the $GW$, named post-$GW$. To obtain Green's function for the post-$GW$, the following steps are used:
\begin{enumerate}
\item Solve the $GW$ equations to get the truncated Green's function $G^{\mathrm{tr}}$, unphysical screened potential $W_{\mathrm{tr}}$, and the Hartree propagator $G^{\mathrm{tr}}_0$.
\item Calculate the covariant vertex using $G^{\mathrm{tr}}$ and $W_{\mathrm{tr}}$, obtaining $\chi^{ab}_{cov}$ \cite{cGW_2023}.
\item Compute the physical screened potential:
\begin{align}
    &W^{ab}_{\mathrm{post}}(1,2)= V^{ab}(1,2)\\
&-\sum_{cd}\int\mathrm d(34)\ V^{ad}(1,4)\chi^{cd}_{cov}(4,3)V^{cb}(3,2).\label{eq:post-W}\nonumber
\end{align}
\item Use $W_{\mathrm{post}}$ to evaluate the post Green's function:
\begin{equation}
\begin{aligned}
\underline G^{-1}_{\mathrm{post}}(1,2)=&
\underline{G_0^{\mathrm{tr}}}^{-1}(1,2)-{\underline\Sigma}^{\mathrm{post}}(1,2),\\
{\underline\Sigma}^{\mathrm{post}}(1,2)=&
-\sum_{ab} \underline{\tau}^{a}\underline{G^{\mathbf{tr}}}\left(1,2\right)\underline{\tau}^{b} W^{ba}_{\mathrm{post}}\left(2,1\right).
\end{aligned}
\label{eq:post-GGW}
\end{equation}
\end{enumerate}
\begin{figure}
\includegraphics[width=0.5\textwidth]{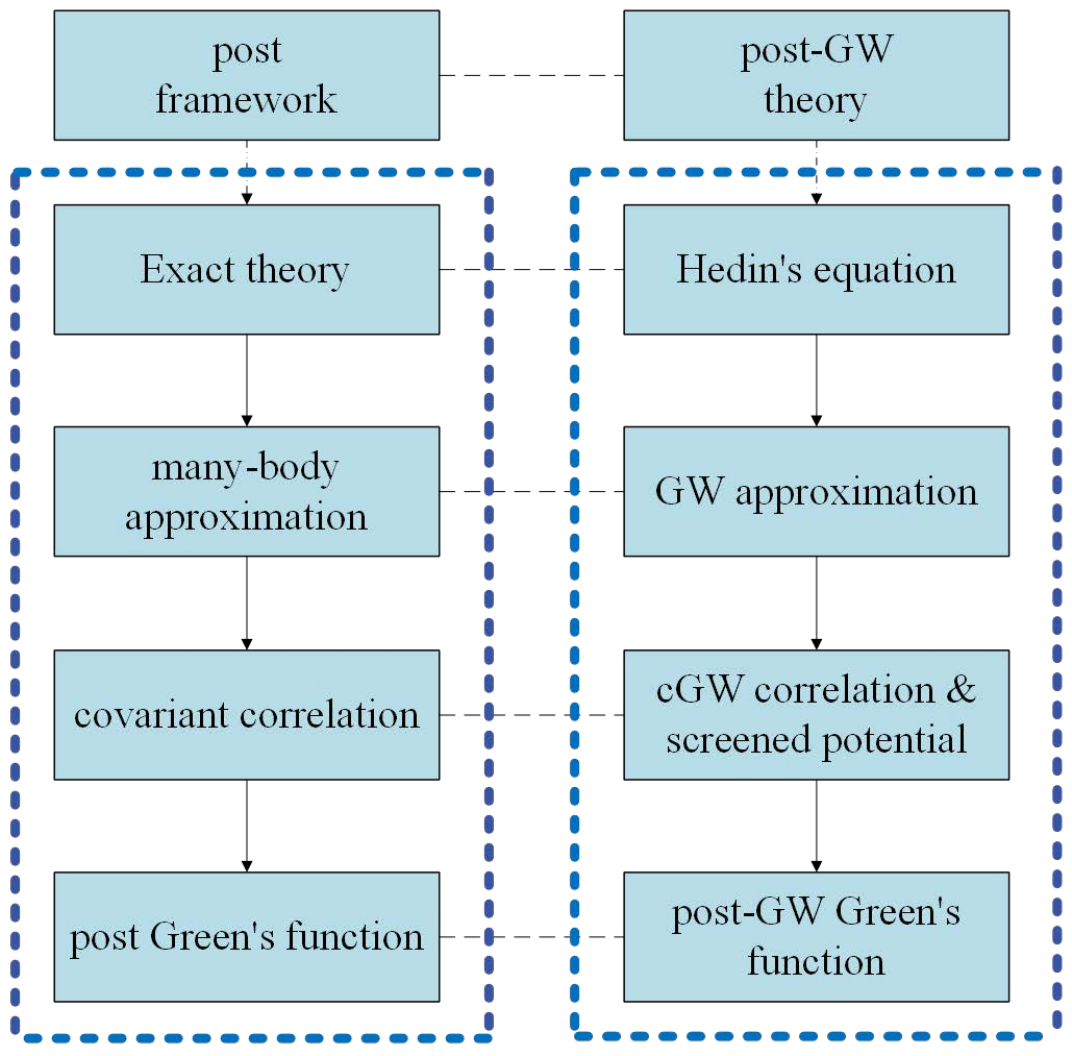}
\captionsetup{justification=raggedright, singlelinecheck=false}
\caption{\label{fig:postgw}The flow chart of the general post framework and the post-$GW$ theory. For the exact theory part, Hedin's equation corresponds to the Eqs.~\ref{eq:postcov-exactG} within the general framework. For the truncated equations part, $GW$ equations correspond to Eqs.~(\ref{eq:postcov-app}). For the covariant part, Eqs.~(\ref{eq:postcov-covarinat}) is represented as Feynman diagrams in Fig.~\ref{FDvertex}, and covariant correlation $\chi^{ab}_{cov}$ and physical screened potential $W_{\mathrm{post}}$ are provided. In the post Green's function calculation part, Eqs.~(\ref{eq:post-GGW}) is the specific form of Eq.~(\ref{eq:postcov-postG}) in the post-$GW$ approach.}
\end{figure}
To demonstrate the relation between the post framework and the post-$GW$ approach, we present the comparison in Fig.(\ref{fig:postgw}).
The Hubbard model is universally acknowledged as a basic model of the strongly correlated systems for high-temperature superconductivity  and is expected to describe pseudogap physics(\cite{qin2022hubbard} and references therein). Therefore, we apply the post-$GW$ approach to the 2D ($L\times L$) single-band Hubbard model and test its performance by directly comparing them with quantum Monte Carlo results.

\textit{Implementation in the Hubbard model.}
---Here the post-$GW$ is used to calculate the Green's function, its spectral function, and the charge compressibility. The Hubbard Hamiltonian with the spin-dependent interaction takes the form \cite{Schafer2021}:
\begin{equation}
    \hat{H}=-\sum_{ij}\sum_{\alpha = \uparrow,\downarrow}t_{ij}\hat{\psi}^{\dagger}_{i\alpha}\hat{\psi}_{j\alpha}- \frac{U}{6}\sum_{i}\vec{\sigma}_{i}\cdot\vec{\sigma}_{i}-\mu\sum_{i\sigma}\hat{n}_{i\sigma}
\end{equation}
where we use the relation $\hat n_{\uparrow}\hat n_{\downarrow} = -\frac{1}{6}\sum_{a = x,y,z} \hat{\sigma^a}\hat{\sigma^a}+\frac{U}{2}\sum_{\alpha}\hat{n}_{\alpha}$ to rewrite the Hubbard interaction form $U\hat n_{\uparrow}\hat n_{\downarrow}$, which can preserve the $SU(2)$ symmetry explicitly in the many-body calculation. Here $\hat\psi^\dagger_{i\alpha}$ creates an electron with spin $\alpha$ at lattice site $i$. $\hat n_{i\sigma}=\psi^{\dagger}_{i\sigma} \psi_{i\sigma}$ denotes the spin-resolved density operator. The hopping amplitude $t_{ij}$ between sites $i$ and $j$ equals $t$ for the nearest neighbors and $t^\prime$ for the next nearest neighbors. Unless otherwise stated, the default values for the hopping amplitude in this paper are $t=1$, $t^\prime=0$. $U$ is the on-site interaction and $\mu$ is the chemical potential.

\begin{figure}
\includegraphics[width=0.8\textwidth]{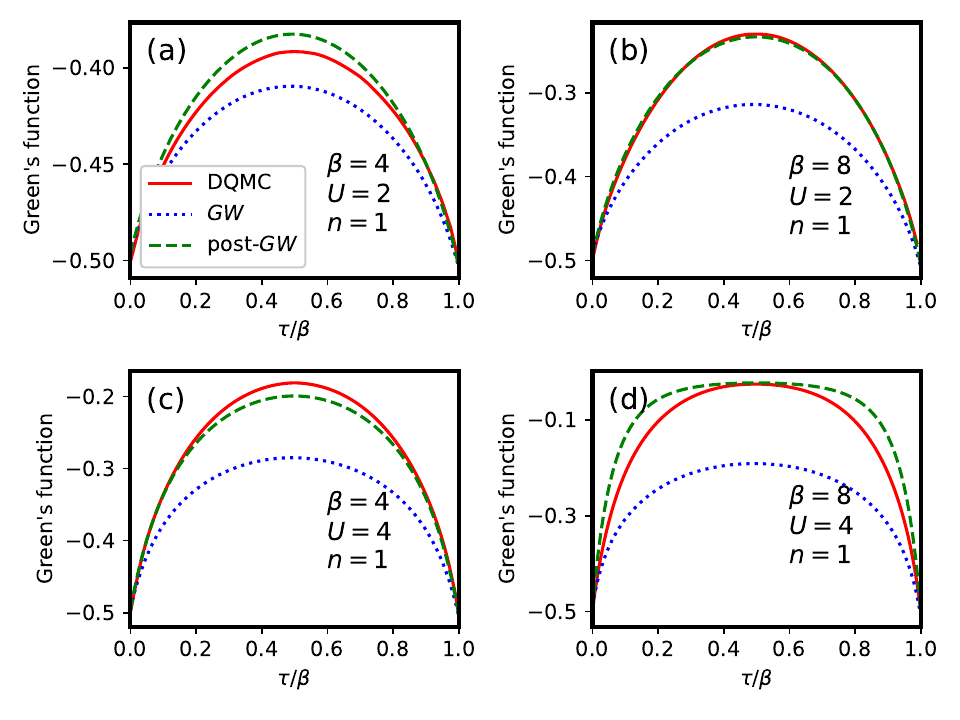}
\captionsetup{justification=raggedright, singlelinecheck=false}
\caption{\label{Green} Green’s functions at the antinodal point $k=(\pi,0)$ in imaginary time space for a $4\times4$ lattice at half-filling for DQMC, $GW$, and post-$GW$ with different parameters:(a)$\beta=4, U=2$,(b)$\beta=8, U=2$,(c)$\beta=4, U=4$,(d)$\beta=8, U=4$.}
\end{figure}

\textit{2D Green's function in the Hubbard model.}---We simulate $4\times 4$ square clusters by $GW$, post-$GW$, and DQMC, which is a well-known numerical exact method at half-filling without the fermion sign problem. Green's functions at the anti-nodal point $\vec k=(\pi,0)$ for the parameters $\beta=4,8$, $U=2,4$ at half-filling in imaginary time are plotted to compare their accuracy directly. Figure~\ref{Green} shows that, for all these parameters, the post-$GW$ approach significantly improves Green’s function, even for the immediate coupling $U=4$, where the traditional $GW$ is regarded as performing poorly.

\begin{figure}
\includegraphics[width=0.8\textwidth]{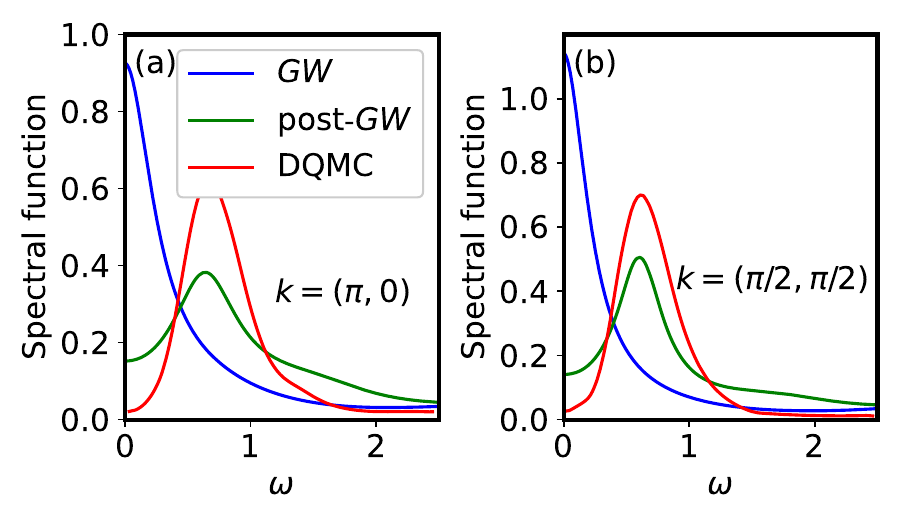}
\captionsetup{justification=raggedright, singlelinecheck=false}
\caption{\label{fig:mott} Comparison of the results of the spectral functions by $GW$, post-$GW$ and DQMC + Maximum Entropy from Ref.~\cite{rost2012momentum} for $8\times8$ cluster at $U=4,T=0.18$, and half-filling at different momenta: (a) $k=(\pi,0)$, (b) $k=(\pi/2,\pi/2)$.}
\end{figure}

To provide robust evidence for the existence of the pseudogap for the immediate $U$ at the low temperature in the post-$GW$ theory, we calculated the Green’s function on an $8\times8$ lattice at $U=4, T=0.18$, and obtained the spectral function by Nevanlinna analytic continuation \cite{fei2021nevanlinna}. 
In Fig.~\ref{fig:mott}, a comparative analysis with the spectral function derived from DQMC with the maximum entropy method as presented in Ref. \cite{rost2012momentum} reveals that, in contrast to the poor results of the $GW$, the post-$GW$ methodology successfully identifies the pseudogap at both the nodal point $k=(\pi,0)$ and antinodal point $k=(\pi/2,\pi/2)$, consistent with the peak positions exposed by DQMC + Maximum Entropy curves. Therefore, the post-$GW$ approach effectively captures the main features of the pseudogap.

To further investigate the performance of the post-$GW$ approach on the two-body correlation functions, we apply the new approach to the calculation of the charge compressibility $\chi=\frac{\partial n}{\partial\mu}$ and compare it with the exact DQMC results \cite{Huang_2019}, where the parameters are taken as $8\times8$ cluster, $t'=-0.25$, $U=6$. The charge compressibility plays an important role in the transport properties of the Hubbard model through the Nernst-Einstein relation, and the doping dependence of the inverse compressibility $\chi^{-1}$ has similar features as the conductivity. Fig.~\ref{fig:invcom} shows that the original $GW$ agrees well with the DQMC in the high-temperature region, but differs from the DQMC qualitatively at low temperatures as the hole doping $p$ decreases. However, the post-$GW$ curves not only are very close to the DQMC curves at high temperatures but also have the same qualitative characteristics as the DQMC at low temperatures with small hole doping $p<0.1$. In particular, in the half-filling case, the DQMC results show that the inverse compressibility has an anomalous increase as the temperature decreases, and post-$GW$ accurately captures such property.

\begin{figure}
\includegraphics[width=0.8\textwidth]{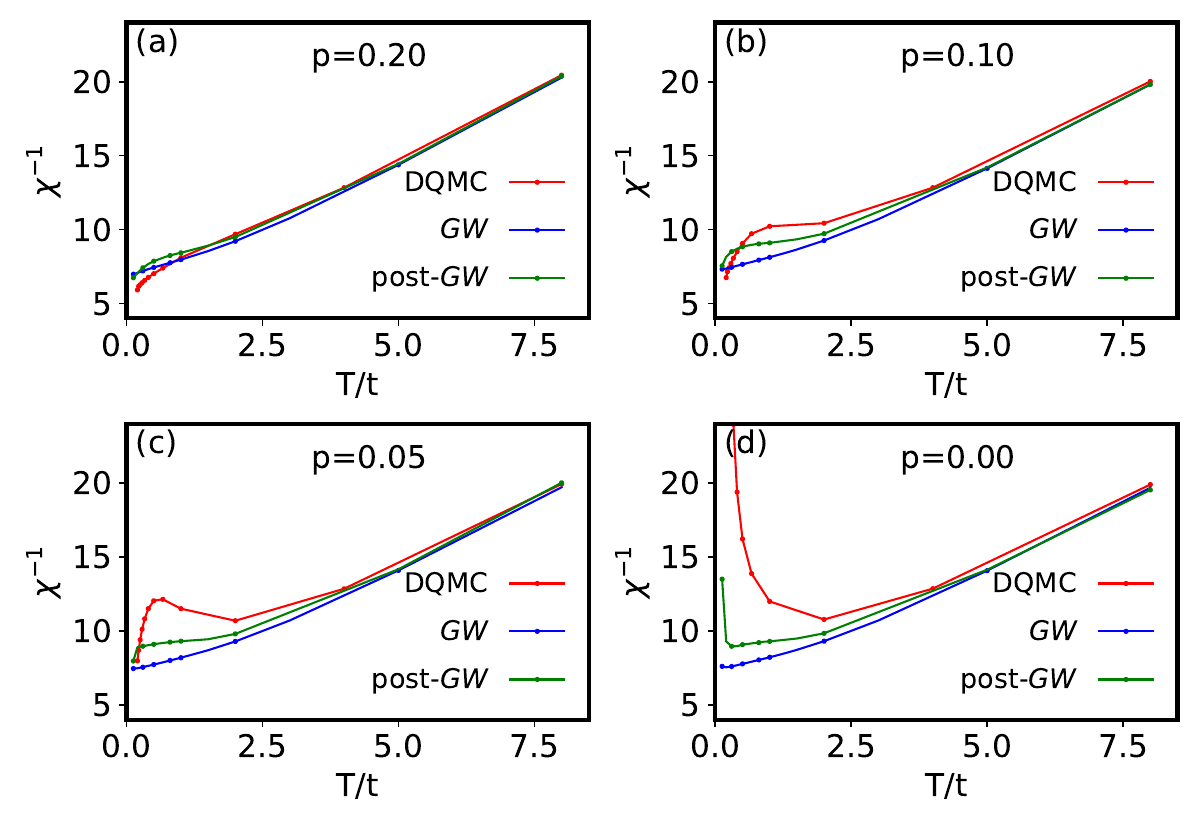}% Here is how to import EPS art
\captionsetup{justification=raggedright, singlelinecheck=false}
\caption{\label{fig:invcom} Inverse charge compressibility $\chi^{-1}$ calculated by $GW$, post-$GW$ and DQMC from Ref.~\cite{Huang_2019} at different dopings. (a) $p=0.2$, (b) $p=0.1$, (c) $p=0.05$, (d) $p=0$, }
\end{figure}

\textit{Conclusions.}---In summary, we propose a theoretical framework for improving Green's function in many-body approximation theories, termed post theory in this article, and apply it to the $GW$ approximation, resulting in the post-$GW$ method. This general framework is based on replacing unphysical higher-order correlations in the self-energy formula with covariant correlations. Specifically in the post-$GW$ approach, we replace the RPA charge/spin correlation in the $GW$ screened potential (\ref{eq:GW-RPAW}) with the corresponding covariant correlation functions.

We apply the post-$GW$ approach to calculate the imaginary-time Green's function, spectral function, and charge compressibility in the 2D Hubbard model. By comparing with the original $GW$ and the exact DQMC results, it is found that post-$GW$ can give significantly better results than $GW$ in both qualitative and quantitative aspects for various physical quantities. In particular, our spectral function calculations reveal that the post-$GW$ approach can capture the pseudogap, which indicates that the $GW$ framework is not incapable of describing the physics of the pseudogap but only needs suitable refinement. We can also apply the post-$GW$ to the 3D Hubbard model, which was recently experimentally studied in the cold atom system \cite{pan2024antiferromagnetic}, and observe the pseudogap phase which is referred to as the paramagnetic Mott state.

In the future, this framework can be expanded to investigate a wide range of correlated systems, including those exhibiting pseudogap phenomena. For example, in the unitary fermion gas, a superconducting pseudogap exists above the phase transition temperature \cite{JinD2008pseudogap, haussmann2009spectral, JinD2010pseudogap,pseudogap2024}. Furthermore, the post framework has the potential to enhance the accuracy of various many-body approximation methods, making it a versatile tool for future research in strongly correlated physics. 

\begin{acknowledgments}
This work is supported by the National Natural Science Foundation of China (Grant No.12174006 of Prof. Li's fund) and the High-performance Computing Platform of Peking University. H.H. acknowledges the support of the National Key R\&D Program of China (No. 2021YFA1401600), and the National Natural Science Foundation of China (Grant No. 12074006 and 12474056). The authors are very grateful to B. Rosenstein, Tianxing Ma, Hong Jiang, Xinguo Ren, Changxiao Li, Wei Wang, and Sheng Yang for valuable discussions and help in numerical computations.
\end{acknowledgments}
\nocite{*}

\providecommand{\noopsort}[1]{}\providecommand{\singleletter}[1]{#1}%

\newpage
\part*{Supplemental Material}
\setcounter{section}{0}
\setcounter{figure}{0}
\setcounter{equation}{0}
\setcounter{page}{1}
\renewcommand{\thesection}{S\arabic{section}}
\renewcommand{\thetable}{S\arabic{table}}
\renewcommand{\thefigure}{S\arabic{figure}}

\section{Hedin's equation and $GW$ approximation for generalized action}
For a generalized action with spin-dependent interaction
\begin{equation}
    S[\psi^*,\psi]=-\sum_{\sigma_1\sigma_2}\int d(12)\psi_{\sigma_1}^*(1)T_{\sigma_1\sigma_2}(1,2)\psi_{\sigma_2}(2)+\frac{1}{2}\int d(12)\sum_{ab}\sigma^a(1)V^{ab}(1,2)\sigma^b(2),
\end{equation}
we consider the perturbation of the system by an external vector source $\vec J(1)$:
\begin{equation}
    S[\psi^*,\psi,\vec J] = S[\psi^*,\psi] - \sum_{a}\int d(1)J^a(1)\sigma^a(1).
\end{equation}
where $\sigma^a(1)=\sum_{\alpha_1\beta_1}\psi^*_{\alpha_1}(1)\tau^a_{\alpha_1\beta_1}\psi_{\beta_1}(1)$, $a,b=0,x,y,z$. Using the grand partition function
\begin{equation}
    Z[\vec J] = \int D[\psi^*,\psi]e^{-S[\psi^*,\psi,\vec J]},
\end{equation}
we can define Green's function and higher-order correlation function
\begin{equation}
    G_{\alpha_1\alpha_2}(1,2) = \left\langle \psi^*_{\alpha_2}(2)\psi_{\alpha_1}(1) \right\rangle,    
\end{equation}
and
\begin{equation}
    G^{(2)}_{\alpha_1\alpha_2\lambda_3\gamma_4}(1,2,3,4) = \left\langle \psi^*_{\alpha_2}(2)\psi_{\alpha_1}(1)\psi^*_{\gamma_4}(4)\psi_{\lambda_3}(3) \right\rangle, \label{G2}\\    
\end{equation}
where $\int D[\psi^*,\psi]$ is the functional integral measure and the ensemble average is defined by $\left< \cdots \right>=\frac{1}{Z}\int D[\psi^*,\psi]\cdots e^{-S}$.

Using the invariance of functional integral measure under the infinitesimal translation transform of the Fermionic fields,
\begin{equation}
    0 = \int D[\psi,\psi^*]\frac{\delta}{\delta \psi^*_{\sigma_1}(1)}(\psi^*_{\sigma_2}(2)e^{-S[\psi^*,\psi;\vec J]}),
\end{equation}
the Dyson - Schwinger equation can be given by:
\begin{eqnarray}
0=&&\delta(1,2)\delta_{\sigma_1,\sigma_2}-\sum_{\sigma_3}\int d(3)T_{\sigma_1\sigma_3}(1,3)G_{\sigma_3\sigma_2}(3,2) \nonumber\\
&&+\sum_{ab}\sum_{\beta_1\lambda_3\gamma_3}\int d(3)\tau^a_{\sigma_1\beta_1}\tau^b_{\lambda_3\gamma_3} V^{ab}(1,3)G^{(2)}_{\beta_1\sigma_2\gamma_3\lambda_3}(1,2,3,3) \nonumber\\
&&-\sum_a\sum_{\beta_1} J^a(1)\tau^a_{\sigma_1\beta_1} G_{\beta_1\sigma_2}(1,2) \nonumber\\
\label{DS1}
\end{eqnarray}
With the definition (\ref{G2}), the two body correlator can be expressed as the derivative of $G$ with respect to $\vec J$:
\begin{equation}
    \frac{\delta G_{\sigma_1\sigma_2}(1,2)}{\delta J^a(3)} 
=\sum_{\alpha_3\beta_3}\tau^a_{\alpha_3\beta_3} G^{(2)}_{\sigma_1\sigma_2\beta_3\alpha_3}(1,2,3,3)-G_{\sigma_1\sigma_2}(1,2)  \left\langle \sigma^a(3) \right\rangle,\label{G2J}
\end{equation}
 Due to the spin structure of the interaction, we denote the matrix in the spin space as
\begin{equation}
    \underline{X}=
    \begin{bmatrix}
    X_{\uparrow\uparrow}& X_{\uparrow\downarrow}\\
    X_{\downarrow\uparrow}&  X_{\downarrow\downarrow}
    \end{bmatrix}.
\end{equation}
Then the Hartree propagator is given by
\begin{equation}
    \underline{H}^{-1}(1,2)=\underline{T}(1,2) +\delta(1,2)\sum_a\underline{\tau}^av^a(1),\label{Hartree}
\end{equation}
and the effective potential $v^a(1)$ is defined by:
\begin{equation}
    v^a(1) = -\int d(4)\sum_{b}V^{ab}(1,4)\left\langle \sigma^b(4) \right\rangle+J^a(1).\label{vD}
\end{equation}
Combine Eqs. ($\ref{DS1}$) and ($\ref{G2J}$), one can obtain:
\begin{equation}
  \delta(1,2)\underline{\tau}^0=\int d(3)\underline{H}^{-1}(1,3)\underline{G}(3,2)-\sum_{ab}\int d(3)\underline{\tau}^a V^{ab}(1,3)\frac{\delta \underline{G}(1,2)}{\delta J^b(3)} \label{DS1_1} 
\end{equation}
For convenience, one can define the vertex function
\begin{equation}
    \underline\Lambda^{a}(1,2,3)=\frac{\delta \underline{G}^{-1}(1,2)}{\delta v^a(3)},\label{Lam}
\end{equation}
and the screened interaction,
\begin{equation}
    W^{ab}(1,2) = \sum_c\int d(3)V^{bc}(2,3)\frac{\delta v^a(1)}{\delta J^c(3)}.\label{WD}
\end{equation}
Notice the identity:
\begin{equation}
    \frac{\delta G}{\delta J} = -G\frac{\delta G^{-1}}{\delta J}G,
\end{equation}
the derivative for Green's function can be expressed as:
\begin{equation}
    \frac{\delta \underline{G}(1,2)}{\delta J^b(3)}=-\sum_c\int  d(456)\underline{G}_{\sigma_1\sigma_4}(1,4) \underline{\Lambda}^c(4,5,6)\underline{G}(5,2)\frac{\delta v^c(6)}{\delta J^b(3)}.\label{GtoJ}
\end{equation}
Then substituting Eq.($\ref{GtoJ}$) to Eq.($\ref{DS1_1}$) leads to:
\begin{equation}
    \delta(1,2)\underline{\tau}^0=\int d(3)\underline{H}^{-1}(1,3)\underline{G}(3,2)+\sum_{ac}\int d(456)\underline{\tau}^a\underline{G}(1,5)\underline\Lambda^c(5,6,4)\underline{G}(6,2)W^{ca}(4,1).
\end{equation}
Combining Eqs.(\ref{vD},\ref{WD}), one arrives at the following equation
\begin{equation}
    W^{ab}(1,2) =\sum_{de}\int d(4567)V^{ad}(1,4)Tr[\underline{\tau}^d\underline{G}(4,6)\underline{\Lambda}^e(6,7,5)W^{eb}(5,2)\underline{G}(7,4)]+V^{ab}(1,2),
\end{equation}
which can be rewritten as:
\begin{equation}
    (W^{-1})^{ab}(1,2) = (V^{-1})^{ab}(1,2)-\int d(34)
 Tr[\underline{\tau}^a\underline{G}(1,3)\underline{\Lambda}^b(3,4,2)\underline{G}(4,1)].
\end{equation}
The inverse of $G,H,W,V$ used above is defined by:
\begin{eqnarray}
&&\int d(2)\underline{G}(1,2)\underline{G}^{-1}(2,3)=\underline{\tau}^0\delta(1,3),\nonumber\\
&&\int d(2)\underline{H}(1,2)\underline{H}^{-1}(2,3)=\underline{\tau}^0\delta(1,3),\nonumber\\
&&\sum_c\int d(3)W^{ac}(1,3)(W^{-1})^{cb}(3,2)=\delta(1,2)\delta^{ab},\nonumber\\
&&\sum_c\int d(3)V^{ac}(1,3)(V^{-1})^{cb}(3,2)=\delta(1,2)\delta^{ab}.
\end{eqnarray}

So, with functional derivative, we can derive the following set of generalized Hedin’s equations:
\begin{align}
    \underline{G}^{-1}(1,2) &= \underline{H}^{-1}(1,2)-\underline{\Sigma}(1,2),\nonumber\\
    \underline{\Sigma}(1,2) &=-\sum_{ac}\int d(45)\underline{\tau}^a \underline{G}(1,5)\underline{\Lambda}^c(5,2,4)W^{ca}(4,1),\nonumber\\
    (W^{-1})^{ab}(1,2) &= (V^{-1})^{ab}(1,2)-P^{ab}(1,2),\nonumber\\
    P^{ab}(1,2) &=\int d(34)Tr[\underline{\tau}^a\underline{G}(1,3)\underline{\Lambda}^b(3,4,2)\underline{G}(4,1)],\nonumber\\
    \underline{H}^{-1}(1,2)&=\underline{T}(1,2) +\delta(1,2)\sum_a\underline{\tau}^av^a(1),\nonumber\\
     v^a(1) &= -\int d(4)\sum_{b}V^{ab}(1,4)Tr[\underline{\tau}^b \underline{G}(4,4)],\nonumber\\
     \underline{\Lambda}^a(1,2,3)&=\frac{\underline{G}^{-1}(1,2)}{v^a(3)}.
\end{align}

The corresponding $GW$ equations can be obtained by simplification of the vertex functions:
\begin{equation}
 \underline{\Lambda}^a(1,2,3) \approx \delta(1,2)\delta(1,3) \underline{\tau}^a.    
\end{equation}
The polarization function and the self-energy then becomes:
\begin{eqnarray}
\underline{\Sigma}(1,2) =-\sum_{ab}\underline{\tau}^a \underline{G}(1,2)\underline{\tau}^bW^{ba}(2,1)\label{GWS}
\end{eqnarray}

\begin{equation}
    P^{ab}(1,2) =
 Tr[\underline{\tau}^a\underline{G}(1,2)\underline{\tau}^b\underline{G}(2,1)].\label{GWP}
\end{equation}

\section{Covariant $GW$ theory}
For the generalized Hartree approximation, which only contains Hartree self-energy in Eq~(\ref{Hartree}), the two-body correlation function obtained by RPA formula can preserve the FDT and WTI (the vertex at the two-body level only contains the first two diagrams in Fig.~\ref{FDvertex}). However, the higher-order approximation, such as $GW$, cannot preserve both the FDT and the WTI when using the RPA formula to calculate the two-body correlation. 

According to the FDT, the two-body correlation functions should be defined as the response of the physical quantity in the presence of an external potential, which we refer to as the covariant scheme. The scheme for calculating a general connected two-body correlation function $\chi_{XY}(1,2)=\left< X(1)Y(2)\right>_c$ within the $GW$ framework, where $X, Y$ are binary composite operators, is formulated as follows.

First, one adds the corresponding source term to the action, $S[\psi^*,\psi;\phi] = S[\psi^*,\psi]-\int d(1)\phi(1) X(1)$ and the correlation can be obtained by $\chi_{XY}(1,2)=\frac{\delta \left< Y(2) \right>}{\delta \phi(1)}$. Then, write down the off-shell $GW$ equations (keep $\phi\neq 0$), and calculate the functional derivative of the $GW$ equations with respect to $\phi$. Finally, let the source $\phi$ tend to $0$ to obtain the on-shell results. Although we restrict our discussion to the $GW$, this scheme can also be applied to different many-body approaches. 

We consider the calculation of a generally connected two-body correlation function
$\chi_{XY}(1,2)=\left\langle X\left(1\right)Y\left(2\right)\right\rangle-\left\langle X\left(1\right)\right\rangle\left\langle Y\left(2\right)\right\rangle$,
where $X,Y$ are local binary operators and take the form
\begin{align}
X\left(1\right) & =\sum_{\alpha_2\alpha_3}\int d(23)\psi^*_{\alpha_2}(2)K_{X;\alpha_2\alpha_3}(1,2,3)\psi_{\alpha_3}(3)\text{.}\label{eq:loc_bin_op}
\end{align}
The expression for the kernel $K$ depends on the operator $X$. As for the spin operator, the kernel $K$ for $\sigma^a(1)=\sum_{\alpha_1\alpha_1^{\prime}}\psi^*_{\alpha_1}(1)\tau^a_{\alpha_1\alpha_1^{\prime}}\psi_{\alpha_1^{\prime}}(1)$ is:
\begin{equation}
    K_{\sigma^a;\alpha_2,\alpha_3}(1,2,3)=\delta(1,2)\delta(1,3)\tau^a_{\alpha_2\alpha_3}.
\end{equation}
First, add an external local source $\phi\left(1\right)$ coupled
to the operator $X\left(1\right)$ and thus the perturbed action becomes
\begin{equation}
S\left[\psi^{\ast},\psi,\phi\right]=S\left[\psi^{\ast},\psi\right]-\int d\left(1\right)\ \phi\left(1\right)X\left(1\right).\label{eq:perturb_action}
\end{equation}
The additional term $\int d\left(1\right)\ \phi\left(1\right)X\left(1\right)$
is explicitly expressed as
\begin{equation}
\int d\left(123\right)\ \sum_{\alpha_{1}\alpha_{2}}\psi_{\alpha_{1}}^{\ast}\left(1\right)\left\{  \phi\left(3\right)K_{X;\alpha_1\alpha_2}(3,1,2)\right\} \psi_{\alpha_{2}}\left(2\right).\label{eq:additional}
\end{equation}
Note that the additional term can be regarded as a variation of the
$T$ term:
\begin{equation}
\underline{T}\left(1,2;\phi\right)=\underline{T}\left(1,2\right)+\int d(3)\phi(3)\underline{K}_{X}(3,1,2).\label{eq:var-T}
\end{equation}

\begin{figure*}
\includegraphics[width=1.0\textwidth,height=0.2\textwidth]{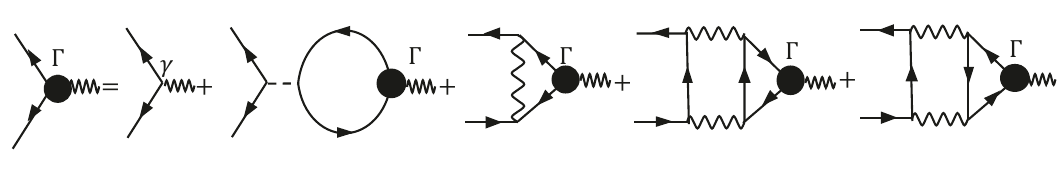}
\captionsetup{font={small},justification=raggedright}
\caption{\label{FDvertex} The Feynman diagram of the full cGW vertex function in Eq.~(\ref{vertex}) for translation invariant systems.}
\end{figure*}

The functional derivative of the off-shell $GW$ equations with respect to the external source $\phi$ leads to the covariant $GW$ (cGW) equations. The equation involves the full vertex function $\underline{\Gamma}_{\phi }(1,2,3)=\frac{\delta \underline{G}^{-1}(1,2)}{\delta\phi(3)}$, which consists of 5 terms shown in Fig.~\ref{FDvertex}:
\begin{eqnarray}
\underline{\Gamma}_{\phi}(1,2,3)=&&\underline{\gamma}_\phi(1,2,3)+\underline{\Gamma}_{\phi,\mathrm{H}}(1,2,3)+\underline{\Gamma}_{\phi,\mathrm{MT}}(1,2,3)\nonumber\\
&&+\underline{\Gamma}_{\phi,\mathrm{AL1}}(1,2,3)+\underline{\Gamma}_{\phi,\mathrm{AL2}}(1,2,3)\label{vertex}.
\end{eqnarray}
Here, the bare vertex $\underline{\gamma}_\phi$ depends on the operator $X$, and is calculated through
\begin{align}
\underline{\gamma}_\phi(1,2,3)&\equiv \frac{\delta \underline{T}(1,2;\phi)}{\delta \phi(3)}\nonumber\\
&=\underline{K}_{X}(3,1,2).\label{eq:bare_vertex}
\end{align}
In the charge/spin response case, $X^a=\sigma^a$, the bare vertex takes the form $\underline{\gamma}_\phi(1,2,3)=\underline{\tau^a}\delta(1,2)\delta(1,3)$. The ``bubble" vertex is induced by the Hartree self-energy, i.e. $\underline{\Gamma}_{\phi,\mathrm{H}}=-\delta\Sigma_{\text{H}}/\delta\phi$, and takes the form:
\begin{equation}
    \underline{\Gamma}_{\phi,\mathrm{H}}(1,2,3) = -\delta(1,2)\sum_{cd}\int d(456)\tau^cV^{cd}(1,4)\text{Tr}[\tau^d \underline{G}(4,5)\underline{\Gamma}_{\phi }(5,6,3)\underline{G}(6,4)].\label{eq:gamma_H}
\end{equation}
Note that the conventional random-phase-approximation-like (RPA) formula only consists of the first two terms in Eq.~(\ref{vertex}).
The Maki-Thompson-like (MT) vertex and two distinct Aslamazov-Larkin-like (AL) vertices are induced by the self-energy, i.e., $\underline{\Gamma}_{\phi}=-\delta\Sigma/\delta\phi$, representing the vertex corrections beyond the RPA, and take the form:
\begin{equation}
    \underline{\Gamma}_{\phi,\mathrm{MT}}(1,2,3)=-\sum_{cd}\int d(45)\underline{\tau}^c\underline{G}(1,4)\underline{\Gamma}_{\phi }(4,5,3)\underline{G}(5,2)\underline{\tau}^d W^{dc}(2,1),\label{eq:gamma_MT}
\end{equation}
\begin{align}
    \underline{\Gamma}_{\phi,\mathrm{AL1}}(1,2,3)&=-\sum_{cdef}\int d(4567)\underline{\tau}^c \underline{G}(1,2)\underline{\tau}^dW^{de}(1,4)W^{fc}(5,2) \nonumber \\
    &\quad \times \text{Tr}\left[\underline{\tau}^e\underline{G}(4,6)\underline{\Gamma}_{\phi }(6,7,3)\underline{G}(7,5)\underline{\tau}^f \underline{G}(5,4)\right],\label{eq:gamma_AL1}
\end{align}
\begin{align}
    \underline{\Gamma}_{\phi,\mathrm{AL2}}(1,2,3)&=-\sum_{cdef}\int d(4567)\underline{\tau}^c \underline{G}(1,2)\underline{\tau}^dW^{de}(1,4)W^{fc}(5,2) \nonumber \\
    &\quad \times \text{Tr}\left[\underline{\tau}^e \underline{G}(4,5)\underline{\tau}^f\underline{G}(5,6)\underline{\Gamma}_{\phi }(6,7,3)\underline{G}(7,4)\right].\label{eq:gamma_AL2}
\end{align}
Finally, let $\phi\to0$ and solve the self-consistent equations (\ref{vertex},\ref{eq:bare_vertex},\ref{eq:gamma_H},\ref{eq:gamma_MT},\ref{eq:gamma_AL1},\ref{eq:gamma_AL2}) to obtain the full vertex function $\underline{\Gamma}_{\phi}$.

Since the average $\left<Y(2)\right>$ is a function of the Green's function $G$, the two-body correlation function $\chi_{XY}(1,2)$ can be obtained by the vertex $\underline{\Gamma}_\phi$:
\begin{align}
\left\langle X\left(1\right)Y\left(2\right)\right\rangle _{\text{c}} & \equiv\frac{\delta\left\langle X\left(1\right)\right\rangle}{\delta\phi\left(2\right)}\nonumber \\
 & =\int d(34)\mathrm{Tr}\left[ \underline{K}_{X}(1,3,4)\frac{\delta\underline G(4,3)}{\delta \phi(2)} \right]\nonumber \\
 & =-\int d(3456)\mathrm{Tr}\left[ \underline{K}_{X}(1,3,4)\underline G(4,5)\underline{\Gamma}_\phi(5,6,2)\underline{G}(6,3) \right].
 \label{eq:tbcf}
\end{align}
For example, when calculating the spin-spin correlation $\chi^{ab}_s(1,2)=\left\langle \sigma^a(1)\sigma^b(2) \right\rangle$, the above equation can be simplified as:
\begin{equation}
    \chi^{ab}_s(1,2)=-\int d(56)\mathrm{Tr}\left[\underline\tau^a\underline G(1,5)\underline{\Gamma}_\phi(5,6,2)\underline{G}(6,1) \right].
\end{equation}
Such response functions satisfy the FDT by definition, and the preserving of the WTI is proven in the next subsection.

\section{Implementation in the two-dimensional Hubbard model}\label{asec:imple}

\subsection{Fourier transformation for a translational invariant lattice}
For a lattice with the translation symmetries, we use the discrete
Fourier transformation to simplify our equations. The Fermionic array
$X_{\text{F}}$ takes the form
\begin{equation}
X_{\text{F}}\left(1,2\right)=\frac{1}{\mathcal{N}}\sum_{k}X_{\text{F}}\left(k\right)\mathcal{E}_{\text{F}}\left(k,1-2\right),\label{eq:f-arr}
\end{equation}
and the Bosonic array $X_{\text{B}}$ takes the form
\begin{equation}
X_{\text{B}}\left(1,2\right)=\frac{1}{\mathcal{N}}\sum_{k}X_{\text{B}}\left(k\right)\mathcal{E}_{\text{B}}\left(k,1-2\right).\label{eq:b-arr}
\end{equation}
Here the transformation kernels $\mathcal{E}_{\text{F}}$ and $\mathcal{E}_{\text{B}}$
are defined as
\begin{equation}
\mathcal{E}_{\text{F}}\left(k,1-2\right)\equiv\text{e}^{\text{i}\vec{k}\cdot\left(\vec{x}_{1}-\vec{x}_{2}\right)}\text{e}^{-\text{i}\frac{2m_{k}+1}{M}\left(l_{1}-l_{2}\right)},\label{eq:f-ker}
\end{equation}
\begin{equation}
\mathcal{E}_{\text{B}}\left(k,1-2\right)\equiv\text{e}^{\text{i}\vec{k}\cdot\left(\vec{x}_{1}-\vec{x}_{2}\right)}\text{e}^{-\text{i}\frac{2m_{k}}{M}\left(l_{1}-l_{2}\right)},\label{eq:b-ker}
\end{equation}
respectively. Here $\mathcal{N}=\beta L^{2}$, $k=\left(\vec{k},m_{k}\right)$
and $m_{k}$ takes the integer value from $0$ to $M-1$. Note that
the transformation of the $T$-term is
\begin{equation}
\underline{T}\left(k\right)=\left[-\frac{1}{\Delta\tau}\left(\text{e}^{-\text{i}\pi\left(2m_{k}+1\right)/M}-1\right)-\varepsilon\left(\vec{k}\right)+\mu\right]\underline{\tau}^{0},\label{eq:t-k}
\end{equation}
with $\varepsilon\left(\vec{k}\right)$ the non-interacting dispersion.
For the two-dimensional Hubbard model, $\varepsilon\left(\vec{k}\right)=-2t\left(\cos k_{x}+\cos k_{y}\right)$
with $t$ the nearest-neighbor hopping strength. 

\subsection{$GW$  and covariant $GW$ equations in Fourier space}

Note that the one-body Green's function $\underline{G}$ and the self-energy
$\underline{\Sigma}$ are Fermionic arrays, and the dynamical potential
$W^{ab}$ and the polarization $P^{ab}$ are Bosonic arrays. It is
easy to derive the $GW$ equations in Fourier space
\begin{align}
\underline{G}^{-1}\left(k\right) & =\underline{T}^{-1}\left(k\right)-\underline{\Sigma}_{\text{H}}\left(k\right)-\underline{\Sigma}\left(k\right),\nonumber \\
\underline{\Sigma}\left(k\right) & =-\frac{1}{\mathcal{N}}\sum_{q,ab}\underline{\tau}^{a}\underline{G}\left(k+q\right)\underline{\tau}^{b}W^{ba}\left(q\right),\nonumber \\
\left(W^{-1}\right)^{ab}\left(q\right) & =\left(V^{-1}\right)^{ab}\left(q\right)-P^{ab}\left(q\right),\nonumber \\
P^{ab}\left(q\right) & =\frac{1}{\mathcal{N}}\sum_{k}\text{Tr}\left[\underline{\tau}^{a}\underline{G}\left(q+k\right)\underline{\tau}^{b}\underline{G}\left(q\right)\right].\label{eq:ggw-fourier}
\end{align}

To derive the covariant $GW$ equations in Fourier space, we first make
ansatz for the vertex function
\begin{equation}
\underline{\Gamma}\left(1,2,3\right)=\frac{1}{\mathcal{N}^{2}}\sum_{p,q}\underline{\Gamma}\left(p,q\right)\mathcal{E}_{\text{F}}\left(k,1-2\right)\mathcal{E}_{\text{B}}\left(q,1-3\right).\label{eq:ansatz-vertex}
\end{equation}
Then one obtains
\begin{equation}
\underline{\Gamma}\left(p,q\right)=\underline{\gamma}\left(p,q\right)+\underline{\Gamma}_{\text{H}}\left(p,q\right)+\underline{\Gamma}_{\text{MT}}\left(p,q\right)+\underline{\Gamma}_{\text{AL1}}\left(p,q\right)+\underline{\Gamma}_{\text{AL2}}\left(p,q\right).\label{eq:gamma-fourier}
\end{equation}
The bare vertex is
\begin{equation}
\underline{\gamma}\left(p,q\right)=\sum_{a}\int d\left(\epsilon\epsilon^{\prime}\right)\ x^{a}\left(\epsilon,\epsilon^{\prime}\right)\underline{\tau}^{a}\text{e}^{\text{i}p\cdot\left(\epsilon-\epsilon^{\prime}\right)}\text{e}^{\text{i}q\cdot\epsilon}.\label{eq:bare-vertex}
\end{equation}
The bubble vertex is
\begin{equation}
\underline{\Gamma}_{\text{H}}\left(p,q\right)=\frac{1}{\mathcal{N}}\sum_{cd}\sum_{k}\underline{\tau}^{c}V^{cd}\left(q\right)\mathrm{Tr}\left[\underline{\tau}^{d}\underline{G}\left(k+q\right)\underline{\Gamma}\left(k,q\right)\underline{G}\left(k\right)\right].\label{eq:bubble-vertex-1}
\end{equation}
The MT vertex is
\begin{equation}
\underline{\Gamma}_{\text{MT}}\left(p,q\right)=-\frac{1}{\mathcal{N}}\sum_{cd}\sum_{k}\underline{\tau}^{c}\underline{G}\left(p+k+q\right)\underline{\Gamma}\left(p+k,q\right)\underline{G}\left(p+k\right)\underline{\tau}^{d}W^{dc}\left(k\right).\label{eq:mt-vertex-1}
\end{equation}
The two AL vertices are
\begin{align}
\Gamma_{\text{AL1}}\left(k,q\right) & =-\frac{1}{\mathcal{N}^{2}}\sum_{cdef}\sum_{kk^{\prime}}\underline{\tau}^{c}\underline{G}\left(p+q+k\right)\underline{\tau}^{d}W^{de}\left(k+q\right)W^{fc}\left(k\right)\nonumber \\
 & \quad\times\mathrm{Tr}\left[\underline{\tau}^{e}\underline{G}\left(k+k^{\prime}+q\right)\underline{\Gamma}\left(k+k^{\prime},q\right)\underline{G}\left(k+k^{\prime}\right)\underline{\tau}^{f}\underline{G}\left(k^{\prime}\right)\right],\label{eq:al-vertex-1-1}
\end{align}
\begin{align}
\Gamma_{\text{AL2}}\left(k,q\right) & =-\frac{1}{\mathcal{N}^{2}}\sum_{cdef}\sum_{kk^{\prime}}\underline{\tau}^{c}\underline{G}\left(p+q+k\right)\underline{\tau}^{d}W^{de}\left(k+q\right)W^{fc}\left(k\right)\nonumber \\
 & \quad\times\text{Tr}\left[\underline{\tau}^{e}\underline{G}\left(k+q+k^{\prime}\right)\underline{\tau}^{f}\underline{G}\left(k^{\prime}+q\right)\underline{\Gamma}\left(k^{\prime},q\right)\underline{G}\left(k^{\prime}\right)\right].\label{eq:al-vertex-2-1}
\end{align}
The diagrammatics for these vertices are presented in Fig.~\ref{FDvertex}.

Note that in the random phase approximation (RPA), the RPA vertex
is given by
\begin{equation}
\underline{\Gamma}_{\text{RPA}}\left(p,q\right)=\underline{\gamma}\left(p,q\right)+\frac{1}{\mathcal{N}}\sum_{cd}\sum_{k}\underline{\tau}^{c}V^{cd}\left(q\right)\mathrm{Tr}\left[\underline{\tau}^{d}\underline{G}\left(k+q\right)\underline{\Gamma}_{\text{RPA}}\left(k,q\right)\underline{G}\left(k\right)\right].\label{eq:rpa-vertex}
\end{equation}
The RPA formula is usually used to calculate the density-density or
spin-spin correlation functions. In the Bethe-Salpeter equation approach,
the MT vertex is taken into account, but the AL vertices are neglected. 

\subsection{$GW$  and covariant $GW$ equations for the 2D Hubbard model}

For the 2D Hubbard model, $\underline{T}\left(k\right)$ takes the
form $T\left(k\right)\underline{\tau}^{0}$ and $V^{ab}\left(k\right)$
takes the form $I^{s}\delta^{ab}$ with $a,b$ taking values of $x,y,z$.
To find the paramagnetic solutions, we can make the ansatz
\begin{equation}
\underline{G}\left(k\right)=G\left(k\right)\underline{\tau}^{0},\ \underline{\Sigma}\left(k\right)=\Sigma\left(k\right)\underline{\tau}^{0},\label{eq:ansatz-1}
\end{equation}
and
\begin{equation}
W^{ab}\left(k\right)=W\left(k\right)\delta^{ab},\ P^{ab}\left(k\right)=P\left(k\right)\delta^{ab}.\label{eq:ansatz-2}
\end{equation}
The $GW$ equation is then simplified as
\begin{align}
G^{-1}\left(k\right) & =T\left(k\right)-\Sigma\left(k\right),\nonumber \\
\Sigma\left(k\right) & =-\frac{3}{\mathcal{N}}\sum_{q}G\left(k+q\right)W\left(q\right),\nonumber \\
W^{-1}\left(q\right) & =1/I^{s}-P\left(q\right),\nonumber \\
P\left(q\right) & = \frac{2}{\mathcal{N}}\sum_{k}G\left(p+k\right)G\left(p\right).\label{eq:ggw-ansatz}
\end{align}

The simplification of the covariant $GW$ equations related to the species
of correlation functions. We take the spin-spin correlation function
as an example here. The spin-spin correlation function $\chi_{\text{s}}^{ab}\left(p\right)$
relates to the vertex function through 
\begin{equation}
\chi_{\text{s}}^{ab}\left(p\right)=-\sum_{q}\text{Tr}\left[\underline{G}\left(p+q\right)\underline{\Gamma}^{a}\left(q,p\right)\underline{G}\left(q\right)\tau^{b}\right].\label{eq:chi-s}
\end{equation}
Here $\underline{\Gamma}^{a}$ refers to the vertex function corresponding
to the spin operator $\sigma^{a}$. By the ansatz $\underline{\Gamma}^{a}\left(q,p\right)=\underline{\tau}^{a}\Gamma\left(q,p\right)$, the spin-spin correlation function $\chi_{\text{s}}^{ab}\left(p\right)=-2\delta^{ab}G\left(p+q\right) \Gamma\left(p,q\right) G\left(q\right)$, and the equation for the vertex function is simplified as
\begin{equation}
\Gamma\left(p,q\right)=\gamma\left(p,q\right)+\Gamma_{\text{H}}\left(p,q\right)+\Gamma_{\text{MT}}\left(p,q\right)+\Gamma_{\text{AL1}}\left(p,q\right)+\Gamma_{\text{AL2}}\left(p,q\right),\label{eq:eq-hub-vertex}
\end{equation}
with the bare vertex $\gamma\left(p,q\right)=1$, the ``bubble''
vertex
\begin{equation}
\Gamma_{\text{H}}\left(p,q\right)=\frac{2I^{s}}{\mathcal{N}}\sum_{k}G\left(k+q\right)\Gamma\left(k,q\right)G\left(k\right),\label{eq:hub-bubble-vertex}
\end{equation}
the MT vertex
\begin{equation}
\Gamma_{\text{MT}}\left(p,q\right)=-\frac{1}{\mathcal{N}}\sum_{k}G\left(p+k+q\right)\Gamma\left(p+k,q\right)G\left(p+k\right)W\left(k\right),\label{eq:hub-mt-vertex}
\end{equation}
and two AL vertices
\begin{equation}
\Gamma_{\text{AL1}}\left(p,q\right)=\frac{2}{\mathcal{N}^{2}}\sum_{kk^{\prime}}G\left(p+q+k\right)W\left(k+q\right)G\left(k+k^{\prime}+q\right)\Gamma\left(k+k^{\prime},q\right)G\left(k+k^{\prime}\right)G\left(k^{\prime}\right)W\left(k\right),\label{eq:hub-al-vertex-1}
\end{equation}
\begin{equation}
\Gamma_{\text{AL2}}\left(p,q\right)=-\frac{2}{\mathcal{N}^{2}}\sum_{kk^{\prime}}G\left(p+q+k\right)W\left(k+q\right)G\left(k+k^{\prime}+q\right)G\left(k^{\prime}+q\right)\Gamma\left(k^{\prime},q\right)G\left(k^{\prime}\right)W\left(k\right).\label{eq:hub-al-vertex-2}
\end{equation}

\subsection{The Details for the Post-$GW$ theory for Fermionic Toy Model}
\subsubsection{Fermionic Toy Model and the Exact solution}
 Our starting point here will be the following “free energy” as a function of the Grassmannian variable $\Psi,\Psi^*$:
 \begin{equation}
S=a(\psi^*_{\uparrow}\psi_{\uparrow}+\psi^*_{\downarrow}\psi_{\downarrow})+b\psi^*_{\uparrow}\psi^*_{\uparrow}\psi_{\downarrow}^*\psi_{\downarrow}.\label{SMeq:toy_model}
 \end{equation}
 Here, $g$ represents “couplings” for the Hubbard-like interaction, and the external source $J$ will be used to calculate correlations and derive Hedin's equation. The exact partition function is:
 \begin{equation}
     Z=\int d \psi_{\uparrow}d\psi^*_{\uparrow}d\psi_{\downarrow}d\psi^*_{\downarrow} e^{-S}
 \end{equation}
 According to the properties of the Grassmannian variables, we can rewrite the exponential term as
\begin{equation}
    e^{-S}=1-a(\psi^*_{\uparrow}\psi_{\uparrow}+\psi^*_{\downarrow}\psi_{\downarrow})-b\psi^*_{\uparrow}\psi_{\uparrow}\psi_{\downarrow}^*\psi_{\downarrow}+a^2(\psi^*_{\uparrow}\psi_{\uparrow}\psi_{\downarrow}^*\psi_{\downarrow}).
\end{equation}
So the partition function can be evaluated exactly:
\begin{equation}
    Z=-b+a^2.
\end{equation}
Here we use the relation:
\begin{align}
    \int d\psi\: \psi &= 1\nonumber\\
    \int d\psi &=0
\end{align}
Similarly, we can obtain the exact Green's function:
\begin{align}
    G=&\left< \psi^*_{\uparrow}\psi_{\uparrow} \right>=\frac{1}{Z}\int D[\psi^*\psi]\psi^*_{\uparrow}\psi_{\uparrow}e^{-S}\nonumber\\
=&\frac{1}{Z}(-a)=-\frac{a}{a^2-b}
\end{align} 
 \subsubsection{Hedin's equation and the $GW$ equations}
  Firstly, we write a more general action
 \begin{equation}
     S[\psi^*,\psi] = -\int d(12)\psi^*(1)T(1,2)\psi(2)+\frac{1}{2}\int d(12)\rho(1)V(1,2)\rho(2),
 \end{equation}
where $1=\alpha$, $\int d(1)=\sum_{\alpha_1}$ $\rho(1)=\psi^*(1)\psi(1)$, and $T(1,2)=-a\delta_{\alpha_1\alpha_2},
V(1,2)=b\delta_{\alpha_1,\bar\alpha_2}$ in the toy model. Therefore, The partition function and the Green's function takes the form:
\begin{equation}
    Z=\int D[\psi^*\psi]e^{-S[\psi^*,\psi]},
\end{equation}
\begin{equation}
    G(1,2)\equiv \left< \psi^*(2)\psi(1) \right>=\frac{1}{Z}\int D[\psi^*\psi]\psi^*(2)\psi(1)e^{-S[\psi^*,\psi]}.
\end{equation}
 To construct the Hedin's equation, we need to introduce the source term in the action:
 \begin{equation}
     S[\psi^*,\psi;J]=S[\psi^*,\psi]-\int d(1)J(1)\psi^*(1)\psi(1).
 \end{equation}
 The invariance of the functional integral measure $D[\psi^*\psi]$ under the infinitesimal variation of field yields:
 \begin{equation}
     \int D[\psi\psi^*]\frac{\delta}{\delta \psi^*(2)}\left( \psi^*(1)e^{-S[\psi^*,\psi;J]} \right).
 \end{equation}
Then one can obtain the Dyson - Schwinger equation
\begin{equation}
    \delta(1,2)=\int d(3)T(1,3)G(3,2)+J(1)G(1,2)-\int d(3)V\left< \psi^*(2)\psi(1)\psi^*(3)\psi(3)  \right>.\label{SMeq:DS-toy}
\end{equation}
One can use
\begin{equation}
    \frac{\delta G(1,2)}{\delta J(3)}=\left< \psi^*(2)\psi(1)\psi^*(3)\psi(3)  \right>-G(1,2)G(3,3)
\end{equation}
to decompose the high-order correlations. Then one can rewrite the Eq.~(\ref{SMeq:DS-toy}) as:
\begin{equation}
    \delta(1,2)
=\int d(3)H^{-1}(1,3)G(3,2)-\int d(3)V(1,3)\frac{\delta G(1,2)}{\delta J(3)}.\label{SMeq:toy-DS1}
\end{equation}
Here the Hartree propagator $H$ is defined as
\begin{equation}
    H^{-1}(1,2)=T(1,2)+\delta(1,2)v(1),\label{SMeq:hartree}
\end{equation}
and the density-weighted effective potential $v$ is:
\begin{equation}
    v(1)=J(1)-\int d(2)V(1,2)G(2,2) \label{SMeq:toy-v}
\end{equation}
To obtain the Hedin's equation, we need to introduce the vertex function
\begin{equation}
    \Lambda(1,2,3)=\frac{\delta G^{-1}(1,2)}{\delta v(3)},\label{SMeq:vertex}
\end{equation}
and the screened potential
\begin{equation}
    W(1,2)\equiv \int d(3)\frac{\delta v(1)}{\delta J(3)}V(2,3). \label{SMeq:toy-W}
\end{equation}
With the definition of the effective potential $v$ and the vertex $\Lambda$, the screened potential can be written in terms of the Green's function and vertex:
\begin{align}
    W(1,2)=&\int d(3)\frac{\delta v(1)}{\delta J(3)}V(2,3)\nonumber\\
    =&V(1,2)-\int d(34)V(1,4)\frac{\delta G(4,4)}{\delta J(3)}V(2,3)\nonumber\\
    =&V(1,2)-\int d(345)V(1,4)\frac{\delta G(4,4)}{\delta v(5)}\frac{\delta v(5)}{\delta J(3)}V(2,3)\nonumber\\
    =&V(1,2)+\int d(4567)V(1,4)G(4,6)\Lambda(6,7,5)G(7,4)W(5,2).
\end{align}

Combine Eq.~(\ref{SMeq:toy-DS1},\ref{SMeq:toy-W},\ref{SMeq:toy-v},\ref{SMeq:hartree},\ref{SMeq:vertex}), one can obtain the Hedin's equation:
\begin{align}
    G^{-1}(1,2)=&H^{-1}(1,2)-\Sigma(1,2),\nonumber\\
    \Sigma(1,2)=&-\int d(34)G(1,3)\Lambda(3,2,4)W(4,1),\nonumber\\
    H^{-1}(1,2)=&T(1,2)+\delta(1,2)v(1),\nonumber\\
    v(1)=&J(1)-\int d(2)V(1,2)G(2,2),\nonumber\\
    W^{-1}(1,2)=&V^{-1}(1,2)-\int d(34)G(1,3)\Lambda(3,4,2) G(4,1).\label{SMeq:toy-Hedin}
\end{align}
One can fine the exact relation between the screened potential $W$ and the high-order correlation $\chi(1,2)=\left< \rho(1)\rho(2) \right>-\left< \rho(1) \right>\left< \rho(2) \right>$:
\begin{align}
       W(1,2)=&\int d(3)\frac{\delta v(1)}{\delta J(3)}V(2,3)\nonumber\\
    =&V(1,2)-\int d(34)V(1,4)\frac{\delta \rho(4)}{\delta J(3)}V(2,3)\nonumber\\
    =&V(1,2)-\int d(34)V(1,4)\chi(4,3)V(2,3).\label{SMeq:chiw}
\end{align}
The lowest-order approximation $\Lambda(1,2,3)\approx\delta(1,2)\delta(1,3)$ would lead to the $GW$  equations, whose self-energy and screened potential take the form:
\begin{equation}
    \Sigma(1,2)=-G^{(tr)}(1,2)W^{(tr)}(2,1),
\end{equation}
\begin{equation}
    (W^{(tr)})^{-1}(1,2)=V^{-1}(1,2)-G^{(tr)}(1,2)G^{(tr)}(2,1)
\end{equation}
Here we use label $(\mathrm{tr})$ to denote that the $GW$ equations are "truncated", not exact. And the $GW$ equations can be explicitly written as:
\begin{align}
    &(G^{\mathrm{tr}})^{-1}_{\alpha_1}=(H^{\mathrm{tr}})^{-1}_{\alpha_1}+G^{\mathrm{tr}}_{\alpha_1}W^{\mathrm{tr}}_{\alpha_1\alpha_1},\nonumber\\
    &(H^{\mathrm{tr}})^{-1}_{\alpha_1}=-a+v_{\alpha_1},\nonumber\\
    &v_{\alpha_1}=\phi_{\alpha_1}-\sum_{\alpha_3}g\delta_{\alpha_1\bar\alpha_3}G_{\alpha_3}^{\mathrm{tr}},\nonumber\\
    &(W^{\mathrm{tr}})^{-1}_{\alpha_1\alpha_2}=1/g\delta_{\alpha_1\bar\alpha_2}-\delta_{\alpha_1\alpha_2}G_{\alpha_1}^{\mathrm{tr}}G_{\alpha_1}^{\mathrm{tr}}.\nonumber
\end{align}

One can notice that the $GW$ equations here can also be applied to lattice systems, like the Hubbard model if the label $(1)$ contains the time, space coordinates, and other freedom.  

\subsubsection{Post-$GW$ Equations for the Toy Model }
According to the post framework, the high-order correlations should be recalculated to be "physical". Here, the only high-order correlation function in the $GW$ equations is density correlation $\chi(1,2)$, which is hidden in the screened potential. So what we should do, is calculate the" physical" $\chi(1,2)$ to obtain the "physical" screened potential $W$.
Here we use the source term $-\int d(1)J(1)\psi^*(1)\psi(1)$ in the original action again. According to our covariant framework, we need to add the source to the free term in the action:
\begin{equation}
    T(1,2;J)=T(1,2)+J(1)\delta(1,2). \label{SMeq:toy-reT}
\end{equation}
Using the new free term $T(1,2;J)$,  we can obtain the $GW$ equations with a non-zero source directly, called off-shell $GW$ equations. Then the functional derivative over the source $J$ in off-shell $GW$ equations would give the covariant two-body vertex with the definition $\Gamma(1,2,3)=\frac{\delta G^{-1}(1,2;J)}{\delta J(3)}|_{J\rightarrow0}$. The covariant equations are:
\begin{align}
    &\Gamma=\gamma+\Gamma_{\mathrm{H}}+\Gamma_{\mathrm{MT}}+\Gamma_{\mathrm{AL}}\label{SMeq:toy-covariant}\\
     &\gamma(1,2,3)=\delta(1,2)\delta(1,3)\nonumber\\
     &\Gamma_{\mathrm{H}}(1,2,3)=-\delta(1,2)\int d(4)V(1,4)\dot G(4,4,3)\nonumber\\
     &\Gamma_{\mathrm{MT}}(1,2,3)=\dot G(1,2,3)W^{\mathrm{tr}}(2,1)\nonumber\\
     &\Gamma_{\mathrm{AL}}(1,2,3)=G^{\mathrm{tr}}(1,2)\dot W(2,1,3)\nonumber\\
     &\dot G(1,2,3)\equiv \frac{\delta G^{\mathrm{tr}}(1,2)}{\delta J(3)}=-\int d(45)G^{\mathrm{tr}}(1,4)\Gamma(4,5,3)G^{\mathrm{tr}}(5,2)\nonumber\\
      &\dot W(1,2,3)\equiv \frac{\delta W^{\mathrm{tr}}(1,2)}{\delta J(3)}=-\int d(45)W^{\mathrm{tr}}(1,4)\Gamma_W(4,5,3)W^{\mathrm{tr}}(5,2)\nonumber\\
      &\Gamma_{W}=-\dot G(1,2,3)G^{\mathrm{tr}}(2,1)-G^{\mathrm{tr}}(1,2)\dot G(2,1,3)\nonumber
\end{align}
Solving these covariant equations, one can obtain the vertex $\Gamma$, and the correlation $\chi(1,2)$ can be calculated through:
\begin{equation}
    \chi_{\mathrm{phy}}(1,2)=\frac{\delta G^{\mathrm{tr}}(1,1)}{\delta J(2)}=-\int d(34)G^{\mathrm{tr}}(1,3)\Gamma(3,4,2)G^{\mathrm{tr}}(4,1).\label{SMeq:toy-chi}
\end{equation}
Such $\chi$ defined by functional derivative is the "physical" correlation function due to the Kubo formula. Then, we use the Eq.~(\ref{SMeq:chiw}) to obtain the renormalized screened potential:
\begin{equation}
W^{\mathrm{post}}(1,2)=V(1,2)-\int d(34)V(1,4)\chi_{\mathrm{phy}}(4,3)V(2,3)
\end{equation}
Finally, we can simulate the Green's function for the post-GW:
\begin{align}
    &G^{-1}_{\mathrm{post}}(1,2)=H^{-1}_{\mathrm{tr}}(1,2)-\Sigma_{\mathrm{post}}(1,2)\nonumber\\
    &\Sigma_{\mathrm{post}}(1,2)=-G^{\mathrm{tr}}(1,2)W^{\mathrm{post}}(2,1)
\end{align}
The key procedure here is using the physical correlation $\chi_{cov}$ to calculate the post-screened potential $W^{\mathrm{post}}$.  

\begin{figure}
\includegraphics[width=0.7\textwidth]{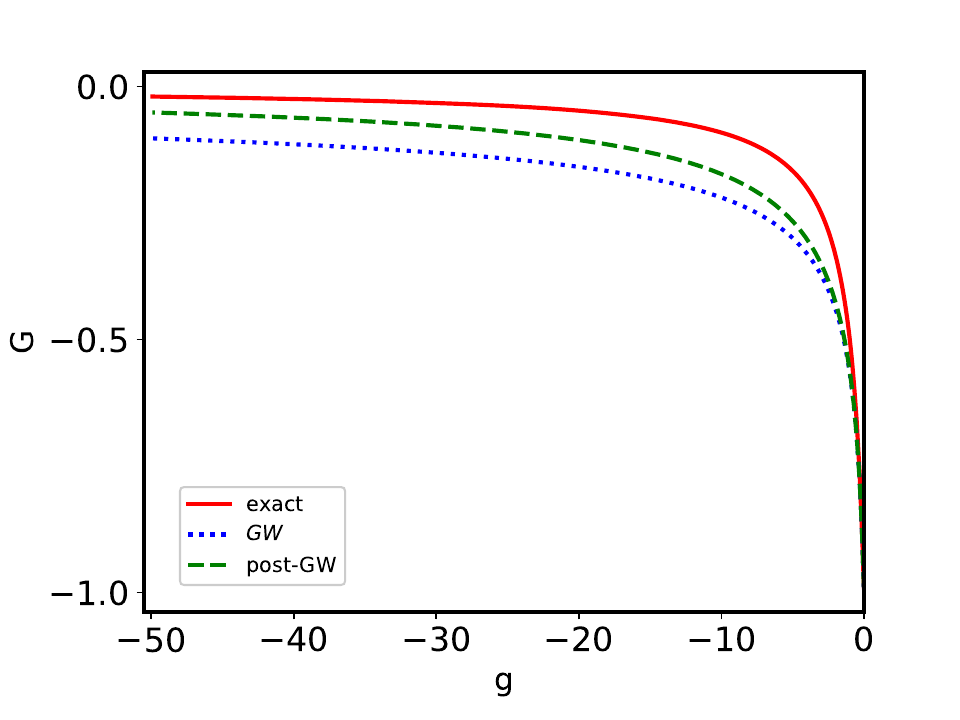}
\caption{\label{fig:post-GW} Green's function obtained by different approaches for an exact solvable model: exact solution (red solid), $GW$ (blue dotted), post-$GW$ (green dashed).}
\end{figure}

We compare the post-$GW$ Green's function $G_{\mathrm{post}}$ with the $G^{\mathrm{exact}}$ and $G^{\mathrm{tr}}$ from the $GW$ theory. Fig.~(\ref{fig:post-GW}) shows that Green's function of the approximate theory is much closer to the exact result after the post correction. The results from this toy model can preliminarily show the validity of the post theory and we will consider a more complex model subsequently.

\section{Benchmark Results for chemical potential dependence of the particle density}
\begin{figure}
\includegraphics[width=0.9\textwidth]{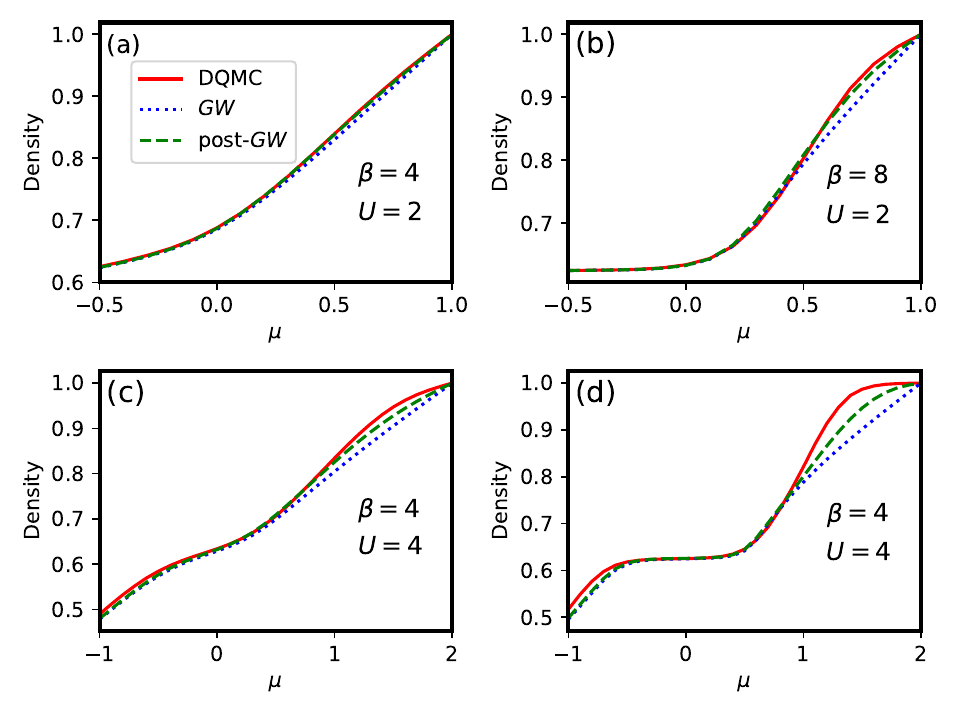}
\caption{\label{fig:Den_vs_mu} The chemical potential dependence of the particle density for DQMC, $GW$ and post-$GW$ at different parameters (a)$\beta=4,U=2$,(b)$\beta=8,U=2$,(c)$\beta=4,U=4$,(d)$\beta=8,U=4$.}
\end{figure}

We study the chemical potential dependence of the particle density obtained from the post-$GW$  Green's function $n=\mathrm{Tr}[\underline{G}(x=0,\tau=0)]$. Fig.~\ref{fig:Den_vs_mu} shows that 
the original $GW$  method deviates substantially from the exact DQMC results when the hole doping $p<0.2$, while the post-$GW$  can significantly correct this deviation in this region. In particular, in Fig.~\ref{fig:Den_vs_mu}(d), where the DQMC result shows a plateau near half-filling indicating a Mott-Hubbard gap phase due to strong antiferromagnetic fluctuation, only post-$GW$  can capture this feature, while $GW$ fails obviously.

\section{Benchmark Results for charge compressitibility}
To provide a more comprehensive analysis of the post-$GW$ approach on the charge compressibility $\chi=\frac{\partial n}{\partial\mu}$, we add more detailed results in Fig.~\ref{fig:invcom}. As shown in Fig.~\ref{fig:invcom}, the original $GW$ method aligns well with the DQMC results at high temperatures but diverges qualitatively at lower temperatures as the hole doping $p$ decreases. In contrast, the post-$GW$ approach not only closely matches the DQMC curves at high temperatures but also maintains similar qualitative features to the DQMC results at low temperatures, particularly when the hole doping $p<0.1$.

\begin{figure*}
\includegraphics[width=0.9\textwidth]{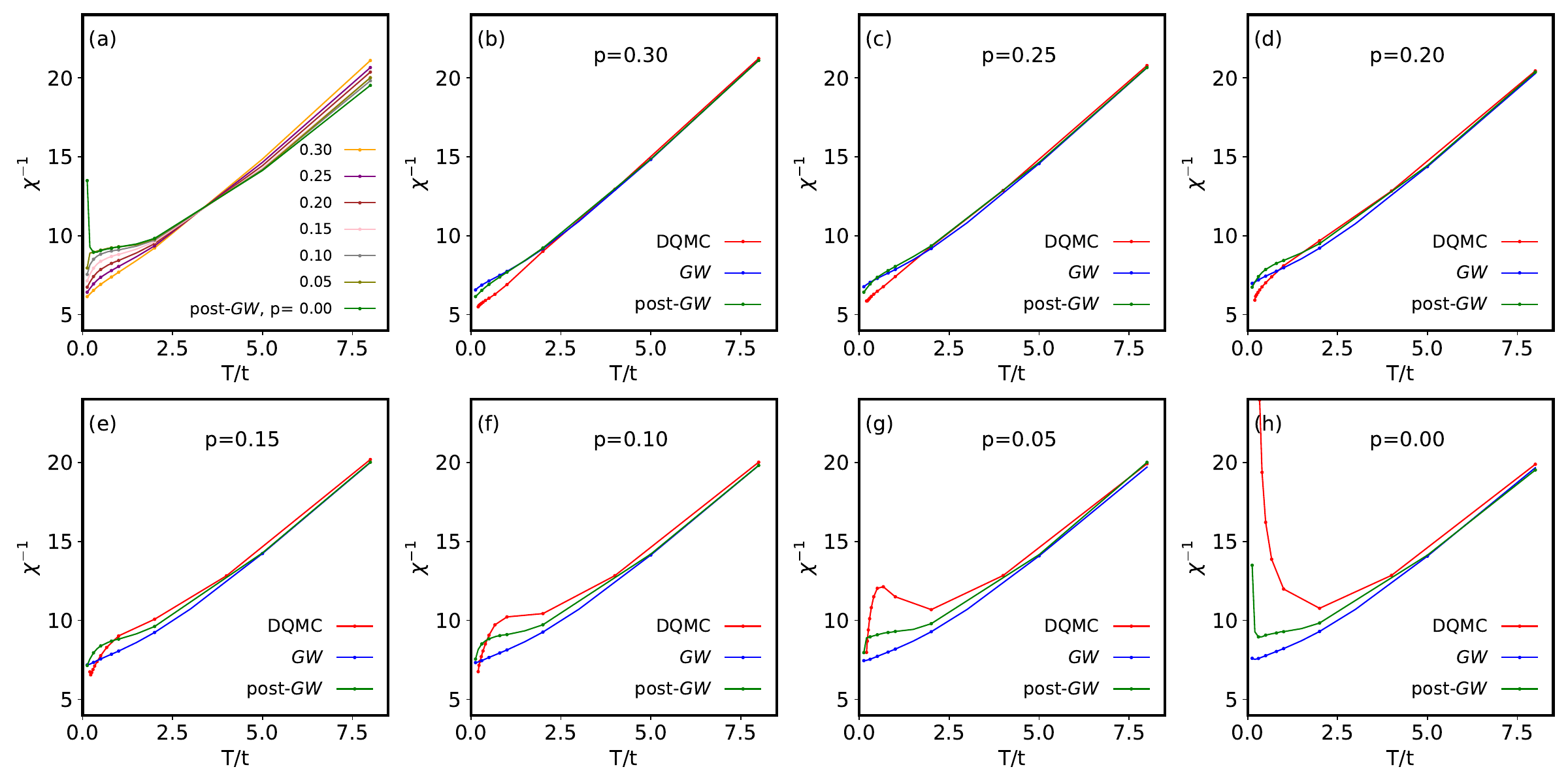}
\captionsetup{font={small},justification=raggedright, singlelinecheck=false}
\caption{\label{fig:invcom} Inverse charge compressibility $\chi^{-1}$ calculated by $GW$, post-$GW$ and DQMC simulations. (a) the curves of the post-$GW$ at different dopings. (b)~(h) The comparisons of the $GW$, post-$GW$, and DQMC at dopings $p=0.3,0.25,0.2,0.15,0.1,0.5,0$.}
\end{figure*}

\end{document}